\documentclass[useAMS,usenatbib]{mn2e}
\usepackage{tabularx}
\usepackage{xspace}
\usepackage{graphicx}
\newcommand{\ignore}[1]{}
\providecommand{\ao}{}

\renewcommand{\ao}{adaptive optics (AO)\renewcommand{\ao}{AO\xspace}\renewcommand{\Ao}{AO\xspace}\xspace}
\newcommand{\Ao}{Adaptive optics (AO)\renewcommand{\ao}{AO\xspace}\renewcommand{\Ao}{AO\xspace}\xspace}
\newcommand{\wfs}{wavefront sensor (WFS)\renewcommand{\wfs}{WFS\xspace}\renewcommand{\wfss}{WFSs\xspace}\xspace}
\newcommand{\wfss}{wavefront sensors (WFSs)\renewcommand{\wfs}{WFS\xspace}\renewcommand{\wfss}{WFSs\xspace}\xspace}
\newcommand{\shwfs}{Shack-Hartmann \wfs (SHWFS)\renewcommand{\shwfs}{SHWFS\xspace}\xspace}
\newcommand{\dm}{deformable mirror (DM)\renewcommand{\dm}{DM\xspace}\renewcommand{\dms}{DMs\xspace}\renewcommand{\Dms}{DMs\xspace}\renewcommand{\Dm}{DM\xspace}\xspace}
\newcommand{\dms}{deformable mirrors (DMs)\renewcommand{\dm}{DM\xspace}\renewcommand{\dms}{DMs\xspace}\renewcommand{\Dms}{DMs\xspace}\renewcommand{\Dm}{DM\xspace}\xspace}
\newcommand{\Dms}{Deformable mirrors (DMs)\renewcommand{\dm}{DM\xspace}\renewcommand{\dms}{DMs\xspace}\renewcommand{\Dms}{DMs\xspace}\renewcommand{\Dm}{DM\xspace}\xspace}
\newcommand{\Dm}{Deformable mirror (DM)\renewcommand{\dm}{DM\xspace}\renewcommand{\dms}{DMs\xspace}\renewcommand{\Dms}{DMs\xspace}\renewcommand{\Dm}{DM\xspace}\xspace}

\newcommand{\lqg}{linear-quadratic-gaussian (LQG)\renewcommand{\lqg}{LQG\xspace}\xspace}
\newcommand{\shs}{Shack-Hartmann sensor (SHS)\renewcommand{\shs}{SHS\xspace}\renewcommand{\shss}{SHSs\xspace}\xspace}
\newcommand{\shss}{Shack-Hartmann sensors (SHSs)\renewcommand{\shs}{SHS\xspace}\renewcommand{\shss}{SHSs\xspace}\xspace}
\newcommand{\lgs}{laser guide star (LGS)\renewcommand{\lgs}{LGS\xspace}\renewcommand{\lgss}{LGSs\xspace}\xspace}
\newcommand{\lgss}{laser guide stars (LGSs)\renewcommand{\lgs}{LGS\xspace}\renewcommand{\lgss}{LGSs\xspace}\xspace}
\newcommand{\Ngs}{Natural guide star (NGS)\renewcommand{\ngs}{NGS\xspace}\renewcommand{\Ngs}{NGS\xspace}\renewcommand{\ngss}{NGSs\xspace}\xspace}
\newcommand{\ngs}{natural guide star (NGS)\renewcommand{\ngs}{NGS\xspace}\renewcommand{\Ngs}{NGS\xspace}\renewcommand{\ngss}{NGSs\xspace}\xspace}
\newcommand{\ngss}{natural guide stars (NGSs)\renewcommand{\ngs}{NGS\xspace}\renewcommand{\Ngs}{NGS\xspace}\renewcommand{\ngss}{NGSs\xspace}\xspace}
\newcommand{\mems}{Micro-Electro-Mechanical Systems (MEMS)\renewcommand{\mems}{MEMS\xspace}\xspace}
\newcommand{\snr}{signal to noise ratio (SNR)\renewcommand{\snr}{SNR\xspace}\xspace}
\newcommand{\Moao}{Multi-object \ao (MOAO)\renewcommand{\moao}{MOAO\xspace}\renewcommand{\Moao}{MOAO\xspace}\xspace}
\newcommand{\moao}{multi-object \ao (MOAO)\renewcommand{\moao}{MOAO\xspace}\renewcommand{\Moao}{MOAO\xspace}\xspace}
\newcommand{\mcao}{multi-conjugate adaptive optics (MCAO)\renewcommand{\mcao}{MCAO\xspace}\xspace}
\newcommand{\ltao}{laser tomographic \ao (LTAO)\renewcommand{\ltao}{LTAO\xspace}\xspace}
\newcommand{\cpu}{central processing unit (CPU)\renewcommand{\cpu}{CPU\xspace}\renewcommand{\cpus}{CPUs\xspace}\xspace}
\newcommand{\cpus}{central processing units (CPUs)\renewcommand{\cpu}{CPU\xspace}\renewcommand{\cpus}{CPUs\xspace}\xspace}
\newcommand{\psf}{point spread function (PSF)\renewcommand{\psf}{PSF\xspace}\renewcommand{\psfs}{PSFs\xspace}\renewcommand{\Psf}{PSF\xspace}\xspace}
\newcommand{\psfs}{point spread functions (PSFs)\renewcommand{\psf}{PSF\xspace}\renewcommand{\psfs}{PSFs\xspace}\renewcommand{\Psf}{PSF\xspace}\xspace}
\newcommand{\Psf}{Point spread function (PSF)\renewcommand{\psf}{PSF\xspace}\renewcommand{\psfs}{PSFs\xspace}\renewcommand{\Psf}{PSF\xspace}\xspace}
\newcommand{\fpga}{field programmable gate array (FPGA)\renewcommand{\fpga}{FPGA\xspace}\renewcommand{\fpgas}{FPGAs\xspace}\xspace}
\newcommand{\fpgas}{field programmable gate arrays (FPGAs)\renewcommand{\fpga}{FPGA\xspace}\renewcommand{\fpgas}{FPGAs\xspace}\xspace}
\newcommand{\sor}{successive over-relaxation (SOR)\renewcommand{\sor}{SOR\xspace}\xspace}
\newcommand{\fdpcg}{Fourier domain pre-conditioned gradient (FDPCG)\renewcommand{\fdpcg}{FDPCG\xspace}\xspace}
\newcommand{\map}{maximum a-posteriori (MAP)\renewcommand{\map}{MAP\xspace}\xspace}

\newcommand{\elt}{Extremely Large Telescope (ELT)\renewcommand{\elt}{ELT\xspace}\renewcommand{\elts}{ELTs\xspace}\renewcommand{\eelt}{European ELT (E-ELT)\renewcommand{\eelt}{E-ELT\xspace}\xspace}\xspace}

\newcommand{\elts}{Extremely Large Telescopes (ELTs)\renewcommand{\elt}{ELT\xspace}\renewcommand{\elts}{ELTs\xspace}\renewcommand{\eelt}{European ELT (E-ELT)\renewcommand{\eelt}{E-ELT\xspace}\xspace}\xspace}

\newcommand{\eelt}{European Extremely Large Telescope (E-ELT)\renewcommand{\eelt}{E-ELT\xspace}\renewcommand{\elt}{ELT\xspace}\renewcommand{\elts}{ELTs\xspace}\xspace}

\newcommand{\dugall}{Durham University generalised adaptive optics laser laboratory (DUGALL)\renewcommand{\dugall}{DUGALL\xspace}\xspace}
\newcommand{\fwhm}{full-width at half-maximum (FWHM)\renewcommand{\fwhm}{FWHM\xspace}\xspace}
\newcommand{\wht}{William Herschel Telescope (WHT)\renewcommand{\wht}{WHT\xspace}\xspace}
\newcommand{\emccd}{electron multiplying CCD (EMCCD)\renewcommand{\emccd}{EMCCD\xspace}\renewcommand{\emccds}{EMCCDs\xspace}\xspace}
\newcommand{\emccds}{electron multiplying CCDs (EMCCDs)\renewcommand{\emccd}{EMCCD\xspace}\renewcommand{\emccds}{EMCCDs\xspace}\xspace}
\newcommand{\dasp}{Durham \ao simulation platform (DASP)\renewcommand{\dasp}{DASP\xspace}\renewcommand{\thedasp}{DASP\xspace}\renewcommand{\Thedasp}{DASP\xspace}\xspace}
\newcommand{\thedasp}{the Durham \ao simulation platform (DASP)\renewcommand{\dasp}{DASP\xspace}\renewcommand{\thedasp}{DASP\xspace}\renewcommand{\Thedasp}{DASP\xspace}\xspace}
\newcommand{\Thedasp}{The Durham \ao simulation platform (DASP)\renewcommand{\dasp}{DASP\xspace}\renewcommand{\thedasp}{DASP\xspace}\renewcommand{\Thedasp}{DASP\xspace}\xspace}
\newcommand{\mpi}{Message Passing Interface (MPI)\renewcommand{\mpi}{MPI\xspace}\xspace}
\newcommand{\smp}{symmetric multi-processing (SMP)\renewcommand{\smp}{SMP\xspace}\xspace}
\newcommand{\svd}{singular value decomposition (SVD)\renewcommand{\svd}{SVD\xspace}\xspace}
\newcommand{\gpu}{graphics processing unit (GPU)\renewcommand{\gpu}{GPU\xspace}\renewcommand{\gpus}{GPUs\xspace}\xspace}
\newcommand{\gpus}{graphics processing units (GPUs)\renewcommand{\gpu}{GPU\xspace}\renewcommand{\gpus}{GPUs\xspace}\xspace}
\newcommand{\fft}{fast Fourier transform (FFT)\renewcommand{\fft}{FFT\xspace}\xspace}
\newcommand{\ifu}{integral field unit (IFU)\renewcommand{\ifu}{IFU\xspace}\xspace}
\newcommand{\darc}{the Durham \ao real-time controller (DARC)\renewcommand{\darc}{DARC\xspace}\renewcommand{\Darc}{DARC\xspace}\xspace}
\newcommand{\Darc}{The Durham \ao real-time controller (DARC)\renewcommand{\darc}{DARC\xspace}\renewcommand{\Darc}{DARC\xspace}\xspace}
\newcommand{\cots}{commercial off-the-shelf (COTS)\renewcommand{\cots}{COTS\xspace}\xspace}
\newcommand{\rtcp}{real-time control pipeline (RTCP)\renewcommand{\rtcp}{RTCP\xspace}\xspace}
\newcommand{\rms}{root-mean-square (RMS)\renewcommand{\rms}{RMS\xspace}\xspace}
\newcommand{\sFPDP}{serial Front Panel Data Port (sFPDP)\renewcommand{\sFPDP}{sFPDP\xspace}\xspace}
\newcommand{\wpu}{wavefront processing unit (WPU)\renewcommand{\wpu}{WPU\xspace}\xspace}

\newcommand{\rtcs}{real-time control system (RTCS)\renewcommand{\rtcs}{RTCS\xspace}\renewcommand{\rtcss}{RTCSs\xspace}\xspace}
\newcommand{\rtcss}{real-time control systems (RTCSs)\renewcommand{\rtcs}{RTCS\xspace}\renewcommand{\rtcss}{RTCSs\xspace}\xspace}
\newcommand{\eso}{European Southern Observatory (ESO)\renewcommand{\eso}{ESO\xspace}\renewcommand{\theeso}{ESO\xspace}\xspace}
\newcommand{\theeso}{\renewcommand{\theeso}{ESO\xspace}the \eso}
\newcommand{\scao}{single conjugate \ao (SCAO)\renewcommand{\scao}{SCAO\xspace}\renewcommand{\Scao}{SCAO\xspace}\xspace}
\newcommand{\Scao}{Single conjugate \ao (SCAO)\renewcommand{\scao}{SCAO\xspace}\renewcommand{\Scao}{SCAO\xspace}\xspace}
\newcommand{\glao}{ground layer \ao (GLAO)\renewcommand{\glao}{GLAO\xspace}\xspace}
\newcommand{\eagle}{ELT Adaptive optics for GaLaxy Evolution (EAGLE)\renewcommand{\eagle}{EAGLE\xspace}\xspace}
\newcommand{\maory}{multi-conjugate \ao relay for the \eelt (MAORY)\renewcommand{\maory}{MAORY\xspace}\xspace}
\newcommand{\muse}{Multi Unit Spectroscopic Explorer (MUSE)\renewcommand{\muse}{MUSE\xspace}\xspace}
\newcommand{\vlt}{Very Large Telescope (VLT)\renewcommand{\vlt}{VLT\xspace}\xspace}

\newcommand{\tmt}{Thirty Metre Telescope (TMT)\renewcommand{\tmt}{TMT\xspace}\xspace}
\newcommand{\xao}{eXtreme \ao (XAO)\renewcommand{\xao}{XAO\xspace}\xspace}

\newcommand{\vla}{Very Large Array (VLA)\renewcommand{\vla}{VLA\xspace}\xspace}
\newcommand{\jwst}{{\em James Webb Space Telescope} \citep[JWST,][]{jwst}\renewcommand{\jwst}{{\em JWST}\xspace}\xspace}
\newcommand{\hst}{{\em Hubble Space Telescope (HST)}\renewcommand{\hst}{{\em HST}\xspace}\xspace}
\newcommand{\ifss}{integral-field spectrographs (IFSs)\renewcommand{\ifss}{IFSs\xspace}\renewcommand{\ifs}{IFS\xspace}\xspace}
\newcommand{\ifs}{integral-field spectrograph (IFS)\renewcommand{\ifss}{IFSs\xspace}\renewcommand{\ifs}{IFS\xspace}\xspace}
\newcommand{\ifus}{integral field units (IFUs)\renewcommand{\ifus}{IFUs\xspace}\xspace}
\newcommand{\mos}{multi-object spectrograph (MOS)\renewcommand{\mos}{MOS\xspace}\xspace}
\newcommand{\goodss}{Great Observatories Origins Deep Survey (GOODS)-S\renewcommand{\goodss}{GOODS-S\xspace}\xspace}
\newcommand{\goods}{Great Observatories Origins Deep Survey (GOODS)\renewcommand{\goods}{GOODS\xspace}\xspace}
\newcommand{\scmos}{scientific CMOS (sCMOS)\renewcommand{\scmos}{sCMOS\xspace}\xspace}
\newcommand{\aof}{Adaptive Optics Facility (AOF)\renewcommand{\aof}{AOF\xspace}\xspace}
\newcommand{\dsp}{digital signal processor (DSP)\renewcommand{\dsp}{DSP\xspace}\renewcommand{\dsps}{DSPs\xspace}\xspace}
\newcommand{\dsps}{digital signal processors (DSPs)\renewcommand{\dsp}{DSP\xspace}\renewcommand{\dsps}{DSPs\xspace}\xspace}
\newcommand{\capi}{Coherent Accelerator Processor Interface (CAPI)\renewcommand{\capi}{CAPI\xspace}\xspace}
\newcommand{\qe}{quantum efficiency (QE)\renewcommand{\qe}{QE\xspace}\xspace}
\newcommand{\numa}{non-uniform memory access (NUMA)\renewcommand{\numa}{NUMA\xspace}\xspace}

\usepackage{amssymb}

\title[Power8 processors for AO real-time control]{Investigation of
  Power8 processors for astronomical adaptive optics
  real-time control}

\author[A.\ G.\ Basden et al.]{A.\ G.\ Basden,$^{1}$\thanks{E-mail:
    a.g.basden@durham.ac.uk (AGB)} \\
$^{1}$Department of Physics, South Road, Durham, DH1 3LE, UK}

\begin{document}
\maketitle
\begin{abstract}
The forthcoming Extremely Large Telescopes all require adaptive optics
systems for their successful operation.  The real-time control for
these systems becomes computationally challenging, in
part limited by the memory bandwidths required for wavefront
reconstruction.  We investigate new POWER8 processor technologies
applied to the problem of real-time control for adaptive optics.
These processors have a large memory bandwidth, and we show that they
are suitable for operation of first-light ELT instrumentation, and
propose some potential real-time control system designs.  A CPU-based
real-time control system significantly reduces complexity, improves
maintainability, and leads to increased longevity for the real-time
control system.
\end{abstract}


\section{Introduction}
The forthcoming \elts \citep{eelt,tmt,gmt} will all rely on \ao systems
\citep{adaptiveoptics} for their successful operation, allowing the
degrading effects of atmospheric turbulence to be greatly reduced.  An
\ao system actively measures wavefront perturbations introduced by the
Earth's atmosphere, and attempts to mitigate these in real-time (on
millisecond timescales) using one or more \dms.  This is
a computationally demanding task, and requires a dedicated \rtcs.
Computational requirements scale with the forth power of telescope
diameter when considering traditional \rtcs algorithms: for a given
level of \ao correction, the \dm pitch must remain constant, and so
the number of sub-apertures across the telescope pupil scales with
telescope diameter, $d$.  The total number of sub-apertures and
actuators therefore each scale as $\mathcal{O}(d^2)$, and therefore
the number of operations required for wavefront reconstruction (a
matrix-vector multiplication) scales as $\mathcal{O}(d^4)$.  Due to
this rapid scaling of computational complexity, careful design
considerations must be made when designing real-time control systems for the \elts.

These \rtcss must be designed with long lifetimes, since the \ao
instruments on these telescopes are expected to be operational for at
least thirty years \citep{30year}.  Therefore maintenance, of both
software and hardware is key to success.  An \rtcs design which is
hardware ambiguous, i.e.\ doesn't require a particular hardware set to
operate, is clearly advantageous.  Previous system designs have
frequently relied on specific hardware, typically \dsps and \fpgas
\citep[for example the ESO SPARTA system,][]{sparta}, which, due to
long periods spent in design, are often close to obsolescence even
during commissioning, with availability of spare parts becoming
problematic, and specific programming knowledge required.  Hardware
failure of these systems then poses the risk that an entire new system
will require designing, with the original software not being portable
to new hardware.

In recent years, there has been much success with hardware agnostic
\ao \rtcss which operate on conventional PC hardware, including \darc
\citep{basden9,basden11}, which is a generic system, used by the
CANARY \ao on-sky demonstrator instrument \citep{canaryshort}, and the
real-time control system for the Gemini South telescope GeMS \ao system
\citep{2012SPIE.8447E..0IRshort}.  In theory, such systems simply
require a recompilation of the source code to be ported to other
(similar) hardware platforms, and are easy to move onto upgraded
hardware.  In practice, the advent of binary driver code, e.g.\ for
\wfss and \dms, means that porting is not always possible.  Although
porting to new hardware is typically limited to other PC-like systems
that have an operating system running on a \cpu, this is not always
the case.  In particular, the \darc system has a modular design which
allows parts of the real-time pipeline to be placed in alternative
hardware, including for example:
\begin{enumerate}
\item pixel processing and slope calculation in \fpga using a
customised version of the SPARTA system \citep{sparta}
\item wavefront reconstruction using \gpus \citep{basden9}
\item a full \gpu pipeline, from raw \wfs images to \dm demands.  
\end{enumerate}
However, this system still requires a \cpu based core to oversee
control of the hardware accelerators.

For \elt-scale \ao systems, the largest computational requirements
come from wavefront reconstruction algorithms, which typically use a
matrix-vector multiplication (MVM) to obtain \dm surface shape
from \wfs slope measurements.  On conventional PC hardware, this
algorithm is memory-bound, rather than compute-bound, and so for low
latency operation, systems with large memory bandwidth are required.
For this reason, accelerator cards (such as \gpus) are considered in
designs for \elt-scale \rtcss to provide the necessary memory bandwidths for
these algorithms.  However, this in itself raises new problems in
moving data into and out of the accelerator for processing, which adds
time and hence latency  to the \rtcs pipeline.  Designs that minimise
this latency are key.

\subsection{The POWER8 processor}
The specification and road-map of the IBM POWER8 processor
\citep{power8short} seems promising for \ao \rtcss, with two key
relevant features: A memory bandwidth approaching that of \gpus (up to
230~GB/s), and support for a novel interconnect technology
\citep[NVLink,][]{nvlink} due for release in 2017 that will provide an
order of magnitude increase in data bandwidth between processor and
\gpu.  Additionally, the OpenPower foundation has the potential for
providing novel hardware acceleration architectures tightly coupled
with POWER8 processors via the \capi \citep{capi}, including a currently available offering from
the company Nallatech.  The memory bandwidth of these processors is
significantly larger than other available \cpus, hence the interest
for \ao real-time control, and a concise overview of the memory
subsystems is given by \citet{power8memory}.

Here, we provide details of initial performance testing of the \darc
\rtcs on a POWER8 system.

In \S2 we discuss the system configuration, \rtcs installation process
and the tests that we perform.  In \S3 we present our findings, and we
conclude in \S4.

\section{The DARC real-time controller on a POWER8 system}
Most of the results that we will present here are performed on a
low-end Tyan OpenPower Customer Reference system, model GN70-BP010,
hosted at Durham.  This system has a single 4-core POWER8 processor
clocked at 3~GHz.  Each core has 8-way symmetric multi-threading, providing a total of
32 hardware threads.  The system has 16~GB DDR3 (1.6~GHz) RAM,
controlled by a single Centaur memory controller.  The total
theoretical memory bandwidth for this system is 28.8~GB/s between \cpu
and main memory (19.2~GB/s read, 9.6~GB/s write).

We have also had limited cloud access to a more powerful S824 POWER8
system with two 12-core processors (to which our machine instance had
access to 22 cores), each 8-way threaded, providing a total of 176
hardware threads.  Half of the memory banks of this machine are
populated, and thus a total memory bandwidth of about 59~GB/s for read
operations, and 29.5~GB/s for write operations is available.  The
operating system of this machine was run behind a hypervisor.  Both of
these systems run the Ubuntu operating system (14.10).  Results
presented here are from our low-end system unless stated otherwise.

\subsection{Real-time control system installation}
We use the publicly available \darc \ao \rtcs system, with source code
downloaded from the {\it sourceforge} hosting site.  Installation on a
POWER8 system was trivial: we simply had to remove three unsupported compiler
options from the Makefile (-msse2 -mfpmath=sse
-march=native) and then compile and install in the usual way.  All of
the required library dependencies were available from the Ubuntu
repositories, and downloaded automatically as part of the \darc
installation process.  We did not attempt to optimise \darc using
compiler flags specific to the POWER8 processor, and we used the
freely available gcc compiler, for which source code is available
(important for lifetime considerations).  

We investigated the use of GigE Vision cameras for wavefront sensors,
using the open-source Aravis library, with modifications specifically
to allow access to the camera pixel stream, rather than full-frame
access (to reduce \rtcs latency).  Because this library is entirely
open-source, and does not require any hardware drivers, there were no
issues with binary drivers.  This library provides access to a number
of wavefront sensors that have been used on-sky with the CANARY \ao
system, including an Imperx Bobcat camera, an
Emergent Vision Technologies HS2000 10GBit camera and a First-Light
OCAM2S camera.  During operation, as soon as sufficient pixels have
arrived at the computer to complete a given sub-aperture, this
sub-aperture is processed by a thread (calibration, slope calculation
and partial reconstruction).  The thread then returns to compute the
next available sub-aperture, in a round-robin fashion.  Once all
sub-apertures for a given frame have been processed, each thread will
have a partial \dm vector, and these are then combined in a reduction
step to yield the final \dm command.

To further demonstrate the proof of concept of a complete \ao system,
we selected an Alpao 241 actuator \dm with an Ethernet interface.  It
was necessary to develop our own library interface for this \dm since
source code for the Software Developers Kit was not available, and the
binary libraries were for X86 architectures.  However, control of this
\dm involves sending a UDP packet, and so was trivial to
implement.  A closed-loop \ao system driven by a POWER8 server is
therefore feasible using an existing \rtcs.

\subsection{Testing real-time performance}

We investigate the performance of \darc on POWER8 by configuring the
system as would be used in a number of different \ao cases.  These
are:
\begin{enumerate}
\item A $40\times40$ sub-aperture \scao system.
\item A $80\times80$ sub-aperture \scao system.
\item A $80\times80$ sub-aperture system with increased actuator counts.
\end{enumerate}

For each of these cases, we investigate performance for different
sized sub-apertures, i.e.\ different numbers of pixels per
sub-aperture.

The third case can be viewed as a single \wfs of the proposed \eelt
\mcao instrument \citep{2010SPIE.7736E..99F} with computation of a full set of
partial \dm demands.  A full \mcao real-time control system
could then be comprised of one compute node per \wfs, with combination
of partial \dm demands being computed as a (low operation count) final
processing step to give the demands to be sent to the \dms.  We
discuss this further in \S\ref{sect:maory}

Our tests presented here do not include a physical \wfs camera or \dm,
since we do not have suitable equipment available (specifically,
cameras with sufficient pixels and frame-rates, and a \dm with enough
actuators).  Rather, we concentrate on the core computational
pipeline.  Our previous experience has shown that introducing a
physical camera to a system has little impact on overall performance
(maximum achievable frame rate), provided the camera
itself is capable of reaching these frame rates.  Because the \darc
\rtcs can process pixels as they arrive at the computer, then once the
last pixel for a given frame arrives, most of the computation has
typically already completed.  The \rtcs is used without frame
pipe-lining here, i.e.\ there are never two frames being processed at
once, so that the frame-rate represents the computation time of a
given frame.  We note that with a real camera, expected readout time
and data transfer time will depend very much on camera model, and in
astronomical \ao the readout time is often the limiting factor in
achievable frame-rate (likely to be the case for the forthcoming
\elts), and for true latency considerations, this should be taken into
account.  For example, for a camera with a maximum frame rate of
500~Hz, the readout time (and exposure time) will be 2~ms.  Assuming
that data is transferred as it is read out (rather than buffered),
this means there will be a delay of 4~ms from start of exposure to
last pixel arriving at the computer (by which time, most of the
computation will have completed).  However, an investigation of camera
latency is beyond the scope of this paper.

Of key importance in the approach that we take is that we are using a fully
configured \ao \rtcs, which has been proven on-sky.
When bench-marking hardware performance, it can be tempting to write
simple bench-marking code which investigates the key algorithms under
consideration, i.e.\ image calibration (vector operations), slope
computation (vector and reduction operations), and wavefront
reconstruction (matrix-vector multiplication).  However, this leads to
optimistic performance estimates, since the bench-mark is grossly
simplified and bears little resemblance to actual code that would be
usable on-sky at a telescope.  

\subsubsection{The performance metric}
We define the performance of the \rtcs by measuring
the time taken to perform the computation for each \ao system frame.
In the default \darc configuration, which we use here, the computation
of each frame must be completed before the next frame is started.
This therefore means that the inverse of the frame computation time
gives the maximum achievable frame-rate for the \ao system.  This
behaviour is critical for optimising \ao system latency on a given
hardware set.  

The \darc \rtcs uses a horizontal processing strategy \citep{basden9}
with each thread operating on \wfs data from start to finish, rather
than having different threads performing individual tasks (e.g.\ a set
of threads for image calibration, a set for slope computation, and a
set for wavefront reconstruction).  This strategy allows automatic
load balancing by the operating system, and simplifies performance
optimisation: the main parameter to be optimised is the number of
processing threads, rather than balancing the number of threads per
algorithm which can become a complex optimisation problem.  Of further
consideration is the number of sub-apertures that each thread should
process at once, influencing the order of memory operations and the
size of the partial matrix-vector multiplications.  If this is too
small, then many inefficient small matrix-vector multiplication
operations will reduce the performance, while if too large, a small
number of large matrix-vector multiplication operations will lead to a
saturation of memory bandwidth, resulting in threads being
work-starved.

\subsection{Tests of memory bandwidth}
To directly test the memory bandwidth available, we use the STREAM
benchmark \citep{streambench}, which performs a number of different
memory read and write operations.  Results are given in
table~\ref{tab:stream}, and show that for our low-end (4-core) server,
over 85\% of theoretical memory bandwidth can be reached, while
achieving nearly 80\% on the higher-end machine.  There are several
things to note here: we did not optimise the STREAM benchmark on the
higher-end machine due to limited access, and so actual performance is
expected to be slightly higher.  The STREAM results
include memory read and write access, which will lead to lower than
expected results for some of these tests since the available
bandwidth on POWER8 systems is asymmetric (i.e.\ the read bandwidth is
twice the write bandwidth).  A non-standard read-only version of
Triad shows slightly higher memory bandwidth utilisation, reaching
90.9\% of the theoretical maximum.

\begin{table}
\begin{tabular}{lll}\hline
STREAM Function & GB/s  & GB/s \\
&(4-core machine) & (22-core machine)\\ \hline
Copy &15.5& 46.0\\
Scale &15.1& 45.5\\
Add &16.3& 41.0\\
Triad &16.4& 46.1\\ \hline
Read-only Triad & 17.4 & \\ \hline

\end{tabular}
\caption{The STREAM benchmark results for the POWER8 systems under
  investigation here (total memory bandwidth achieved).  For the
  4-core machine, best performance was using 3 threads, while 48
  threads were used for the 22-core machine.  The Read-only line is an
  additional function that we added to test read memory access only
  (i.e.\ no memory writes are performed), and is achieved using 4
  threads.}
\label{tab:stream}
\end{table}

\section{RTCS performance on POWER8}
We now consider the achievable performance on the POWER8 systems under
investigation, and consider the application for future \rtcs designs.
For each case, we investigate changing the number of threads used by
\darc, and the processing block size used, i.e.\ the number of
sub-apertures processed together as a block.

\subsection{An 8~m XAO system}
\begin{figure}
\includegraphics[width=\linewidth]{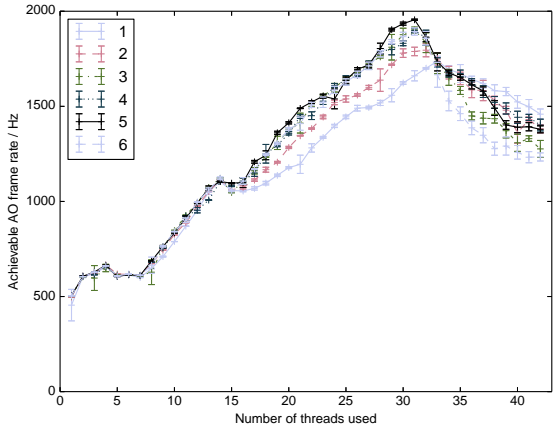}
\caption{Achievable RTCS frame rate as a function of number of
  processing threads used.  The individual lines represent the number
  of times (given by the legend) threads are reused each frame
  (affecting the number of partial matrix-vector products that are
  implemented).}
\label{fig:scao40}
\end{figure}

We investigate the case of a \xao system on an 8~m telescope with
20~cm sub-apertures ($40\times40$), and results are shown in
Fig.~\ref{fig:scao40}.  Here, it can be seen that with the low-end
system a maximum frame-rate of nearly 2~kHz is achieved.  In this case, the
control matrix size is $1304\times2480$, requiring a memory bandwidth
of 23.4~GB/s to read this from main memory every \rtcs iteration at
this frame rate.  This is larger than the available memory bandwidth
(19.2~GB/s) and therefore, the control matrix (12~MB) is being stored in the
large L3 cache (32~MB).  

\rtcs processing tasks are divided among a selected number of threads,
and we see that using 31 threads provides best performance.  The
processor has 4 cores, each with 8-way simultaneous multithreading
capability (i.e.\ 32 virtual cores).  Of particular note is the
linearity of these curves between 8 threads and the peak: the \rtcs
pipeline is seen to be highly parallelisable with performance scaling
almost directly with the number of cores available.

\ignore{
The higher-end system provides a maximum frame-rate of about 850~Hz
(requiring 10~GB/s memory bandwidth to retrieve the control matrix
from main memory every iteration). We do not understand why this
result is worse than on our low-end system, and have not had enough
access to the system to investigate. xxx
}

\begin{figure}
\includegraphics[width=\linewidth]{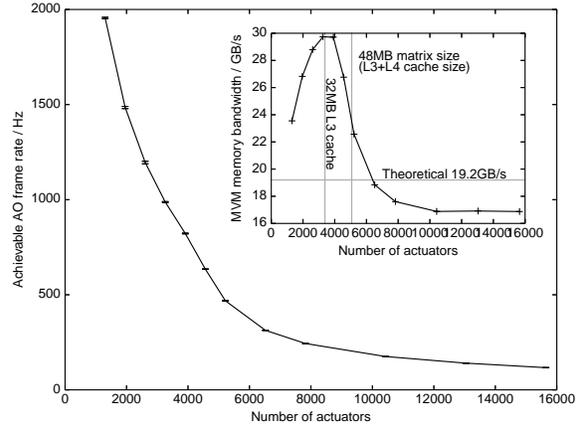}
\caption{Maximum achievable RTCS frame rate as a function of number of
actuators controlled for a $40\times40$ sub-aperture system.  Inset is
shown the corresponding memory bandwidth required by the matrix-vector
multiplication to achieve this frame rate.  }
\label{fig:scao40nact}
\end{figure}

We also consider the case when this system has a larger number of
actuators to control, e.g.\ for a woofer-tweeter system.  This is of
particular interest, because it will allow us to measure maximum \rtcs
performance as the control matrix size approaches, and exceeds, that
of the L3 cache.  Fig.~\ref{fig:scao40nact} shows these results (with
the optimum number of processing threads selected), which shows an
expected degradation of achievable \ao frame rate as the problem size increases.
Once the control matrix size approaches about 48~MB (equal to the size of the
L3 and L4 cache combined), then performance is clearly
degraded, with memory bandwidth between the processor and main memory
becoming the limiting factor.  Performance levels off
utilising about 90\% of the available memory bandwidth for large
control matrix sizes, in agreement with the STREAM benchmark.

\subsection{A single ELT WFS}
We investigate the case of an \eelt \scao system, with a single \wfs
with $80\times80$ sub-apertures (with $6\times6$ pixels per
sub-aperture), and a control matrix of size $5160\times9824$ (193~MB).
In this case, the maximum frame rate is 100.2~Hz on our low-end
system, requiring a memory bandwidth of 18.9~GB/s to read the matrix
from memory each iteration (it is too large to fit in cache), in
addition to reading calibration image and other memory operations.
This is very close to the theoretical maximum memory bandwidth, and so
we conclude that the POWER8 architecture is optimised and pipelined in
such a way as to achieve peak performance for mixed processing tasks.

The higher-end system provides a maximum frame-rate of 150~Hz,
requiring a memory bandwidth of 28.8~GB/s (with a slightly larger
control matrix with 10,000 actuators).  It should be noted that
because of the way the \rtcs is currently implemented, a single copy
of the control matrix is accessed, and therefore will be stored in the
memory attached to one processor.  Threads executing on the second
processor must therefore access this matrix via the first processor,
therefore limiting the available memory bandwidth for control matrix
access to that of one processor, i.e. 29.5~GB/s in this case.  This is
clearly a limiting factor for the \rtcs, in part due to the \numa
architecture of the multi-processor computer hardware, one which is
now on the list of improvements to be made to the \darc system.  We note
here that we are achieving an effective memory bandwidth very close to
the theoretical limit available to the system.

For reference a top-end Intel X86 processor (E5-2699-v3) has 18 cores
and a 45~MB level-3 cache, with 68~GB/s access to main system memory,
costing around £5000.

\begin{figure}
\includegraphics[width=\linewidth]{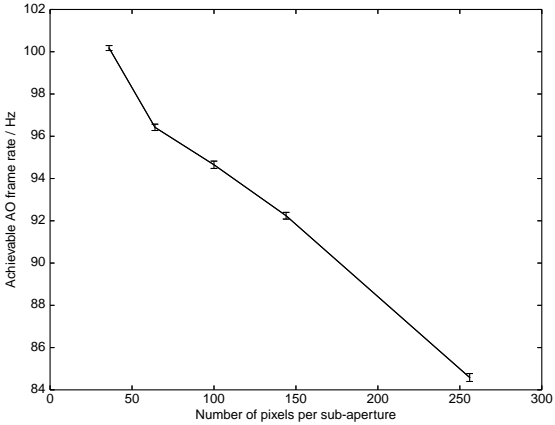}
\caption{Maximum AO frame rate as a function of number of pixels
  per sub-aperture (with $80\times80$ square sub-apertures).}
\label{fig:perf}
\end{figure}

We also investigate the effect of number of pixels on \ao real-time
performance, with Fig.~\ref{fig:perf} showing maximum \ao frame
rate on our low-end POWER8 hardware as a function of number of pixels
per sub-aperture.  Increasing the number of pixels per sub-aperture
reduces maximum frame-rate, suggesting that as sub-apertures get
larger, the matrix-vector multiplication is no longer the sole rate
limiting factor.  Although the memory bandwidth required to read an
image, background map and flat-field information at the \ao frame rate
is small (compared to that required for the control matrix), at only
1.5~GB/s for the largest sub-apertures used here, the larger images
will have a larger impact on cache operations, meaning that less of
the control matrix is available in cache for when required, leading to
additional memory reads, and reduced \ao frame-rates.  Additionally, a
larger number of floating point operations are required for pixel
processing, meaning that the matrix-vector multiplication time is no
longer so dominant.

\subsubsection{Thread counts}
\begin{figure}
\includegraphics[width=\linewidth]{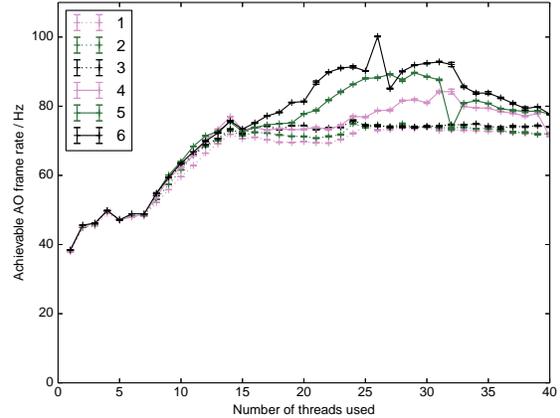}
\caption{A figure showing how maximum achievable AO frame rate is
  dependent on the number of processing threads used.  The individual
  lines represent the number of times (given by the legend) threads
  are reused each frame.}
\label{fig:nthreads}
\end{figure}

We investigate how the number of processing threads affects the
achievable \ao frame rate.  Fig.~\ref{fig:nthreads} shows that using close
to, but less than, the number of hardware threads (32) provides best
performance.  Of particular note here is that (in comparison with
Fig.~\ref{fig:scao40}) performance no longer scales directly with the
number of processing cores.  This is because this larger problem size
is memory bandwidth limited, rather than compute limited.

\subsubsection{Amdahl's law}
\label{sect:nomvm}
Amdahl's law \citep{amdahl} states that the performance gain in a
system through parallelisation (or other) techniques is limited by the
fraction of time spent within the parts of the system benefiting from
those improvements.

In the case of a high order \ao \rtcs, the limiting performance factor
is memory bandwidth, required for wavefront reconstruction.
Increasing available memory bandwidth will only continue to
significantly improve performance while other parts of the
computational pipeline (namely image calibration and slope
calculation) do not begin to dominate the computation time.
Therefore, to be able to make scaled performance predictions, we need
to be able to determine the time taken for these operations which are
compute limited rather than memory bandwidth limited.

We therefore investigate performance with and without wavefront
reconstruction.  For the case without wavefront reconstruction, we are
interested in how well the POWER8 system can process pixel information
and produce wavefront slopes, and assume that the reconstruction could
be performed elsewhere (i.e.\ in a \gpu, using NVLink), though of
course this may introduce additional latency.  

\begin{figure}
\includegraphics[width=\linewidth]{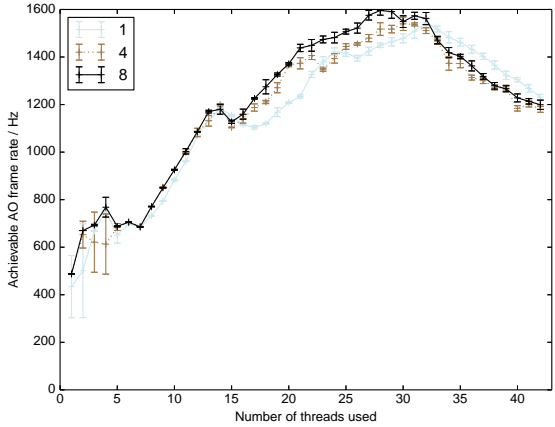}
\caption{A figure showing achievable AO RTCS frame-rates as a function
  of thread count on the low-end POWER8 system when wavefront
  reconstruction is not performed, for an ELT-scale SCAO system
  ($80\times80$ sub-apertures).}
\label{fig:nomvm}
\end{figure}

Fig.~\ref{fig:nomvm} shows maximum achievable frame rates for the \ao
\rtcs processing pipeline when the large matrix-vector multiplication
for wavefront reconstruction is removed, and thus places an
approximate limit on achievable performance for these processors when
unlimited memory bandwidth is available.  Therefore, we can see that
when using a POWER8 system with greater memory bandwidth (up to
256~GB/s read bandwidth for a dual-processor server), frame rates of
nearly 1.3~kHz should be available for this system, limited by the
memory bandwidth for wavefront reconstruction, since we know that
other aspects of the real-time pipeline can be performed faster than
this (1.6~kHz on our low-end system, and faster on a high end 24-core server).

\subsection{A multiple mirror ELT SCAO system}
To investigate the performance of this \elt-scale \scao system further, we
consider the case of multiple mirror \scao systems, i.e.\ with an
increased number of actuators.  This increases the control matrix
size, and thus allows us to investigate performance limiting factors
for different \ao system configurations.  We also investigate
performance with different sub-aperture sizes (pixels per
sub-aperture), so that we can separate compute intensive and memory
intensive tasks.

\begin{figure}
\includegraphics[width=\linewidth]{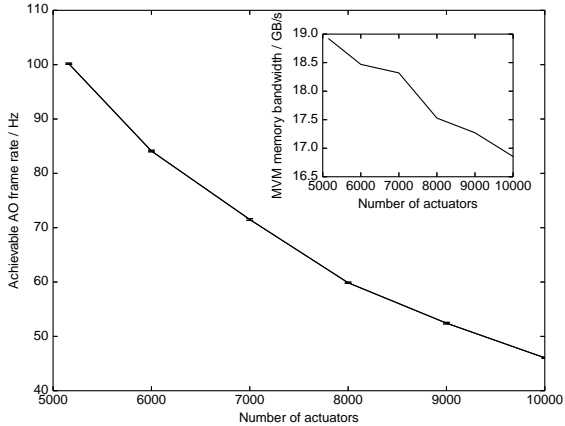}
\caption{Maximum AO frame rate as a function of number of actuators
  controlled with $80\times80$ sub-apertures.  Inset is shown the
  memory bandwidth required reach this frame rate for a given matrix
  size.  }
\label{fig:perfsize}
\end{figure}

Fig.~\ref{fig:perfsize} shows maximum \ao frame rate on our low-end POWER8
hardware as a function of control matrix size.

The maximum achievable frame-rate is reduced proportionally to the
control matrix size, again limited by memory bandwidth, though we see
that for larger matrices, the memory bandwidth achieved is slightly
reduced.  We believe that this is due to less of the larger matrix
being cached, i.e.\ when there is a larger matrix to read, cache
prediction is not so good.  However, the system is still able to
achieve nearly 90\% of theoretical memory bandwidth during the \ao
system loop.

\ignore{1.5GB/s:  (80*16)**2 *4bytesPerPxl * framerate * 3(1img,1ff,1bg) /2**30.
6  100.175  0.117287
8  96.4227  0.150469
10 94.6511  0.173491
12 92.2399  0.158849
16 84.5842  0.187381

10000  46.0569  0.119722
9000  52.4344	0.153022
8000  59.8674	0.126609
7000  71.5115	0.163347
6000  84.1127	0.155964
5160  100.175   0.117287

Bandwidth for 10k, 9k, 8k,7k,6k,5160:
[ 16.85555971,  17.2705945 ,  17.52785854,  18.31987065,  18.46976202,  18.91722929]

9824 slopes, 4912 subaps.
}
\subsubsection{Operation at necessary frame rates}
The maximum frame rates reported so far have not been sufficient for
an on-sky \elt \ao system.  However, we have only been able to perform
bench marking on a low-end system.  Due to the high utilisation of
available memory bandwidth (close to 100\%), we can make predictions
as to maximum achievable frame rates for currently available higher
end systems.  A POWER8 S824 system contains two processors, each with
up to 128~GB/s memory bandwidth for read operations, a combined factor
of 13.3 times greater than our system.  If memory bandwidth is the
limiting factor, we could expect an \ao frame rate of greater than
1.2~kHz for an \elt-scale \scao system using an S824 system.  It is
likely that other parts of the computational pipeline would start to
limit performance so that this frame rate would not be achieved.  In
\S\ref{sect:nomvm} we have investigated performance on our low-end
system with the matrix-vector multiplication removed, to demonstrate
that pixel processing and slope computation at higher frame rates is
achievable.  Therefore, with sufficient memory bandwidth, \elt frame
rates are easily available on an existing POWER8 server.

\subsection{An ELT MCAO system}
\label{sect:maory}
We have considered the performance case for an \elt-scale \scao
system, and we now use this information to consider \mcao system
design.  The \eelt \mcao instrument, MAORY
\citep{2010SPIE.7736E..99F}, is likely to have 4--6 \lgss and up to 3
\ngs low order wavefront sensors, with a total of 2 or 3 \dms
(including the telescope M4 \dm), operating up to 10,000 actuators
with a 500~Hz frame rate.

Processing of \wfs images to yield wavefront gradients is independent,
i.e.\ slopes obtained by processing one \wfs do not depend on the
processing of other \wfss.  Similarly, when using conventional
matrix-vector multiplication wavefront reconstruction methods (we
discuss other methods in \S\ref{sect:lqg}, the slopes from each \wfs
can be used independently of other \wfss to compute a partial set of
\dm commands.  The partial \dm commands from each \wfs can then be
summed, yielding the final \dm demands to be applied to the mirror, in
a low count vector addition operation.

\begin{figure}
\includegraphics[width=\linewidth]{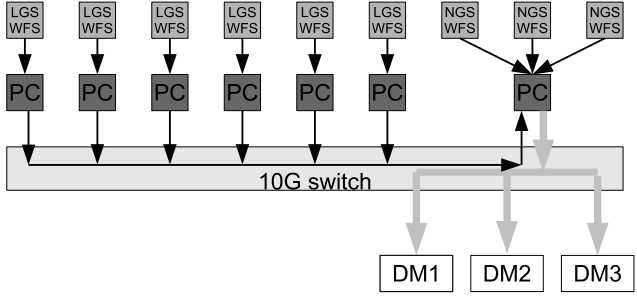}
\caption{A schematic design showing components for a ELT MCAO
  real-time control system, and the links between them.  WFSs are
  connected individually to a POWER8 server, which computes partial DM
demands.  These are then summed before being sent to the DM.}
\label{fig:maory}
\end{figure}

We therefore now consider a \mcao control solution which has a separate
POWER8 server for each \lgs \wfs (directly connected), and an
additional POWER8 server for the three \ngs, with partial \dm
demands being sent to one server for summation to yield the final \dm
demands, as shown in Fig.~\ref{fig:maory}.  We note that since the
\ngs are likely to be of lower order (resulting in a smaller
matrix-vector multiplication), it would be possible to process all
\ngs in a single server, reducing cost and complexity.  This server is
then also used to collate the partial \dm demands, which will arrive
over more than one 10G Ethernet link to reduce latency.

With this control solution, each server therefore has to process a
single \wfs, and between 8000--10000 actuators, and so we can directly
estimate expected performance using Fig.~\ref{fig:perf}, which by
scaling to the memory bandwidth available in a S824 system, will yield
frame rates above 500~Hz, the MAORY design goal.  Further processor
improvements over the next few years (for example the Power9 processor
in 2017) will improve performance further, and be available within the
time frame of MAORY system development.

\subsection{An ELT MOAO system}
We now consider requirements for an \elt-scale \moao system.  The
\eelt \moao instrument is likely to be MOSAIC \citep{Hammer2014}, and will
use 6 \lgs and up to 5 \ngs.  Up to 20 \moao channels are proposed, each
with a \dm, in addition to the main telescope M4 deformable mirror.  

\begin{figure}
\includegraphics[width=\linewidth]{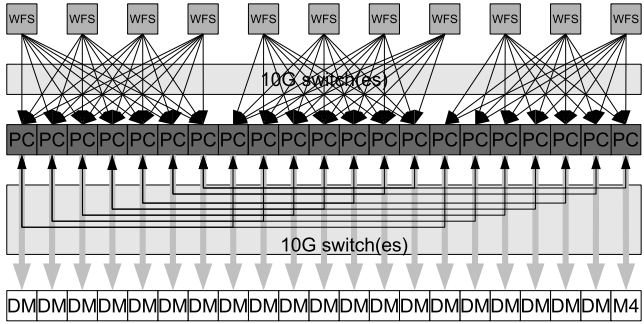}
\caption{A schematic design showing components for a ELT MOAO
  real-time control system, and the links between them.  Four WFSs are
connected to a server, which computes slope measurements, and shares
these with two other servers.  Each server then has access to all
wavefront sensor slope measurements, and computes DM demands for a
single DM.}
\label{fig:mosaic}
\end{figure}

Fig.~\ref{fig:mosaic} shows a possible schematic design for the \moao
real-time control system.  In summary, 21 servers are required, one
for each \dm, including the M4 mirror.  Each server receives images
from 3 or 4 \wfss and processes these to provide wavefront slope
information.  These wavefront slopes are then shared with two other
servers, which in return also share the wavefront slope information
computed from their \wfss.  Therefore, each server will have access to
the 11 \wfs slope vectors.  Each server then performs a tomographic wavefront
reconstruction, projected along a given line of sight, and sends the
\dm demands to the relevant \dm.  

With this design, each server is responsible for processing 4 \wfs
images, and performing a matrix-vector multiplication with a matrix
size of about $100,000\times5000$.  At the desired frame rate of
250~Hz, this represents a required memory bandwidth of about 470~GB/s,
which is achievable using a 4-socket POWER8 server (e.g.\ the S850
system, which has a read memory bandwidth of 512~GB/s), though is
above that obtainable in a single dual socket server.
It is likely that within the next decade (the time-frame for \elt \moao
instrument development), significant improvements in
memory bandwidth will be realised, enabling this performance goal to
be met with even greater overhead, reducing latency.  Additionally,
the inclusion of one or two
\gpus to the system \citep[taking advantage of the forthcoming high
performance NVLINK interconnect,][]{nvlink} specifically to perform
matrix-vector multiplication would further reduce latency.  We discuss this
further in \S\ref{sect:lqg}.  

It should be noted that with this design, the wavefront reconstruction
for each \dm is independent, allowing different algorithms to be
trialled with performance comparisons made while the system is in
operation.  This capability will be key to maximising \moao performance.

\subsection{Variation in latency}
\begin{figure}
\includegraphics[width=\linewidth]{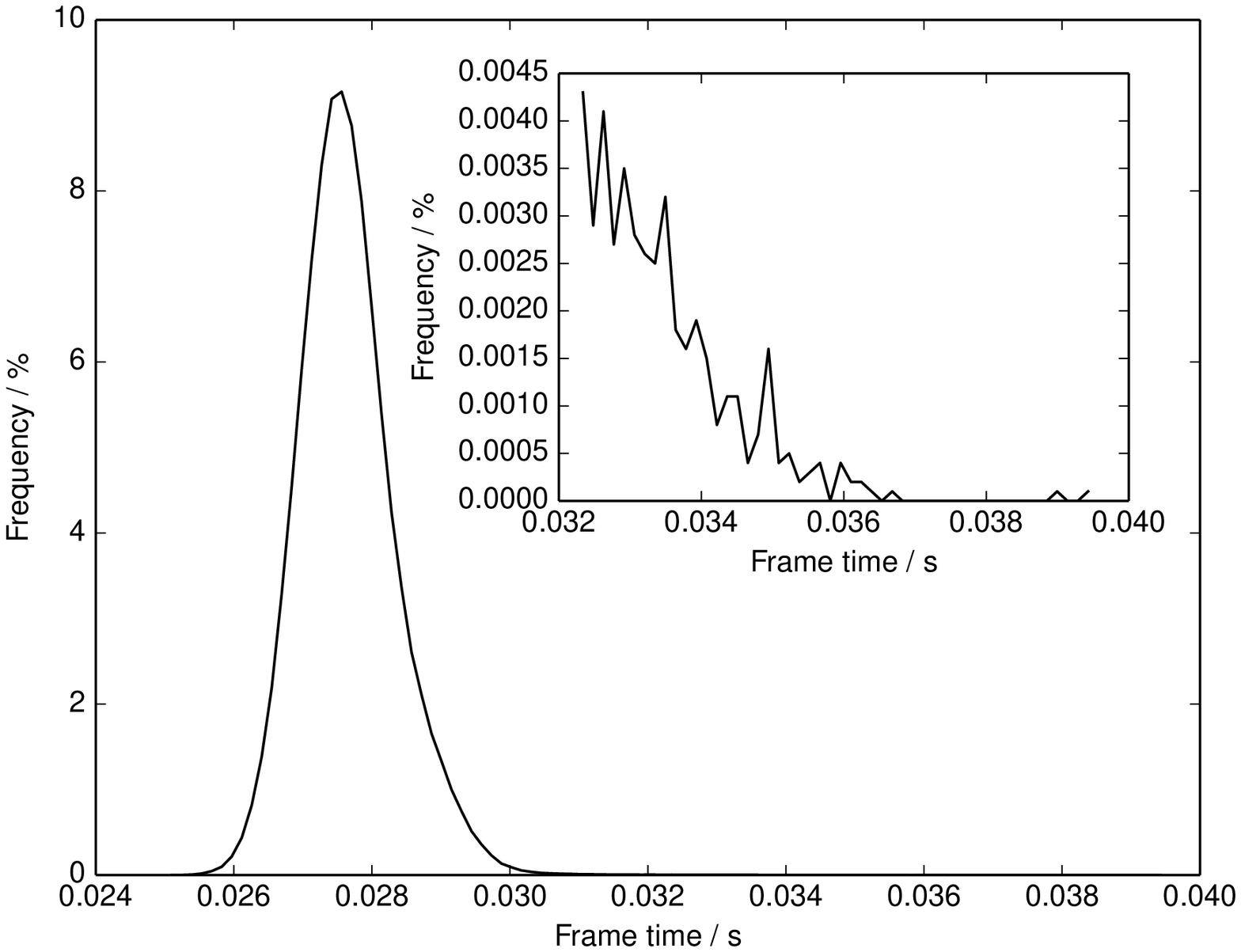}
\includegraphics[width=\linewidth]{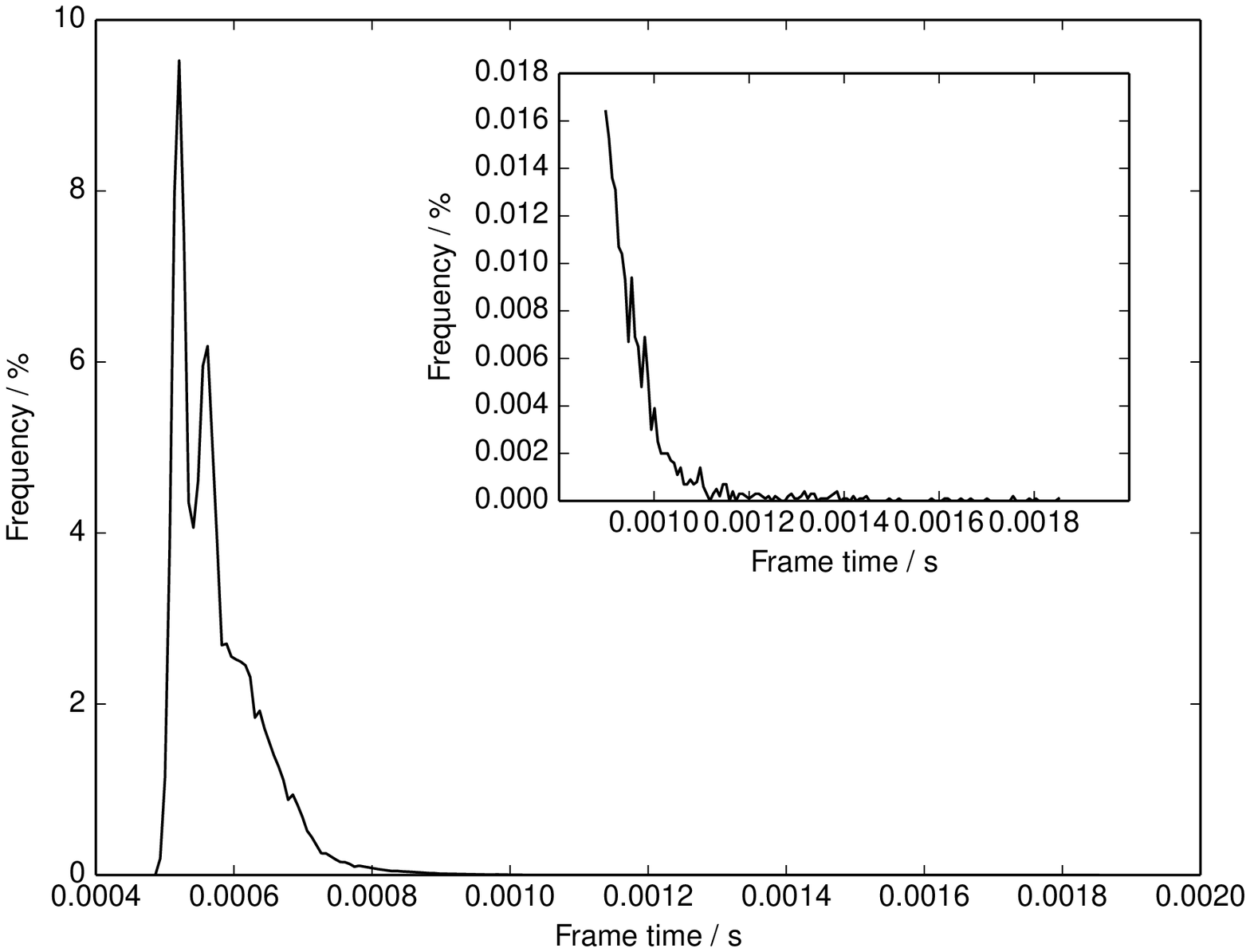}
\caption{(a) A histogram of frame computation times for an
  $80\times80$ SCAO system.  (b) A histogram of frame computation
  times for a $40\times40$ SCAO system.}
\label{fig:jitter}
\end{figure}

The variation of \ao system latency, or jitter, is a key parameter
when developing a real-time control system.  If this jitter is large,
then there will be frequent delays in the \ao processing pipeline,
leading to reduced \ao performance.  This is particularly critical for
higher order \ao systems.  Fig.~\ref{fig:jitter} shows the variation
in latency measured over 1,000,000 frames on the POWER8 server for
both the $40\times40$ and $80\times80$ sub-aperture systems.  For the
higher order case, the variation in latency follows a Gaussian
distribution, with a FWHM of 1.4~ms, 5\% of the mean frame time.  No
frames take more than twice the mean frame time, and 99\% of frames
take less than 8\% longer than the mean time.
\ignore{Stats:
>>>hist=numpy.histogram(data,100)
>>>hist[0][:31].sum()
989250
>>>hist[0][:32].sum()
992833
>>>((hist[1][1:]+hist[1][:-1])/2.)[32]
0.029732399992644785
Mean time: 0.02764859629  - so ratio of these is 1.0754 (8\%).
}

For the low order case, the variation in latency is no longer
Gaussian, showing an extended tail, and additional features that may
be related to the granularity of the timer.  The rms jitter is
62$\mu$s.  Here, less than 0.01\% of frames take longer than twice the mean
frame time to complete, and 99\% of frames take less than 38\% longer
than the mean frame time to complete.

\ignore{
>>> hist2[0][:43].sum()
989358
{71}>>> hist2[0][:44].sum()
990444
{72}>>> x2[44]  #x2=(hist2[1][1:]+hist2[1][:-1])/2.
0.00078827028162777426
>>>x2[44]/data2.mean()
1.3712328937551608

}

We are currently using a stock Ubuntu kernel (3.16.0-23).  The use of
a real-time kernel would further improve this jitter, though we do not
investigate here as this is not yet available.

\subsection{Further considerations}
\label{sect:lqg}
We have so far only considered the basic \ao \rtcs pipeline
operations, including wavefront reconstruction using a matrix-vector
multiplication algorithm, image calibration and slope computation.
However, for an \elt, this is unlikely to be sufficient, as further
algorithms will be necessary, for example the \lqg control as
demonstrated by CANARY, for vibration mitigation \citep{lqgshort}, which
involves several matrix-vector multiplication operations.  

Current implementations of \lqg demonstrated on-sky have required
significantly more computational power and memory bandwidth than a
conventional matrix-vector multiplication algorithm, and so the
hardware that we are investigating here may not be sufficient for
these algorithms.  There are two alternatives: \lqg is an active area
of research, and efficient implementations are being developed
\citep{efficientlqg}.  Alternatively, hardware acceleration techniques
can be considered.

A requirement for additional hardware acceleration will benefit
significantly from the proposed high-speed NVLINK and \capi
interconnects under development for future POWER8
processors and hardware accelerators.  Specific hardware, for example
\gpus or \fpgas, can be used to provide acceleration of given
algorithms, in this case, the wavefront reconstruction problem.  A
high-speed, low latency link is key to enabling this, as it will
maintain low system latency: improved algorithmic behaviour will only
improve \ao system performance if the algorithms do not lead to
significant increases in \ao system latency.  A key feature of the
\capi interface is that it enables abstracted code to be developed
with accelerators sharing the same memory address space as the \cpu,
allowing code to be developed independently of the physical hardware
acceleration used.

A high bandwidth, low latency accelerator interconnect is also
essential for future designs of \elt-scale \xao real-time systems.
For these systems, low latency is critical.

\subsubsection{Future-proofing AO real-time control}
We have demonstrated that an existing \ao \rtcs can be ported to
an alternative processor technology with very little effort, and that
this technology has the potential to enable \ao real-time control for
first-light \elt \ao instruments without the requirement for additional
hardware acceleration.  This greatly simplifies \rtcs design, and
provides greater confidence that the \rtcs software will be able to
operate for the foreseeable future, independent of underlying hardware
changes (provided a C compiler exists).  No proprietary libraries are
necessary, and full source code for this system is available.  

Of key importance here is that an \elt-scale \ao real-time control
system can be developed in the widely used C programming language, and
does not require any custom hardware, or any niche untransferable
skills.  Transferability of this system to other processor types give
a significant degree of confidence that a system developed in this way
will remain operable, configurable, upgradable and hardware
independent for the foreseeable future.  This is a key advantage for
telescopes with expected operational lifetimes approaching a century.

\section{Conclusion}
We have investigated the use of a freely available, open source, \ao
\rtcs on new POWER8 hardware.  We find that installation on this
hardware was trivial, demonstrated the use of \wfss and a \dm, and
find that computational performance is in line with expectations, with
\elt-scale \ao \rtcs performance being limited by available memory
bandwidth, of which our \rtcs typically reaches above 90\% of the
theoretical maximum.  The large potential memory bandwidth of the
POWER8 \cpu, along with forthcoming innovations enabling high
bandwidth communication between the \cpu and other hardware (including
\gpus, with NVLink), means that POWER8 systems are a prime contender
for use with \elt-scale \ao \rtcss, and that using conventional
computer server technology is highly attractive to maintain longevity,
upgradability and comprehension of these systems.

\section*{Acknowledgements}
This work is funded by the UK Science and Technology Facilities
Council, grant ST/K003569/1 and ST/L00075X/1.  We thank the referee who helped improve
this paper.

\bibliographystyle{mn2e}

\bibliography{mybib}

\begin{thebibliography}{}

\bibitem[\protect\citeauthoryear{{Amdahl}}{{Amdahl}}{1967}]{amdahl}
{Amdahl} G.,  1967, in {AFIPS Conference Proceedings, Volume 30, pp. 483-485
  (1967)} {Validity of the Single Processor Approach to Achieving Large-Scale
  Computing Capabilities}.
pp 483--485

\bibitem[\protect\citeauthoryear{{Babcock}}{{Babcock}}{1953}]{adaptiveoptics}
{Babcock} H.~W.,  1953, \pasp, 65, 229

\bibitem[\protect\citeauthoryear{{Basden}, {Geng}, {Myers} \&
  {Younger}}{{Basden} et~al.}{2010}]{basden9}
{Basden} A.,  {Geng} D.,  {Myers} R.,    {Younger} E.,  2010, Appl.\ Optics,
  49, 6354

\bibitem[\protect\citeauthoryear{{Basden} \& {Myers}}{{Basden} \&
  {Myers}}{2012}]{basden11}
{Basden} A.~G.,  {Myers} R.~M.,  2012, \mnras, 424, 1483

\bibitem[\protect\citeauthoryear{{Fedrigo}, {Donaldson}, {Soenke}, {Myers},
  {Goodsell}, {Geng}, {Saunter} \& {Dipper}}{{Fedrigo} et~al.}{2006}]{sparta}
{Fedrigo} E.,  {Donaldson} R.,  {Soenke} C.,  {Myers} R.,  {Goodsell} S.,
  {Geng} D.,  {Saunter} C.,    {Dipper} N.,  2006, in Advances in Adaptive
  Optics II. Edited by Ellerbroek, Brent L.; Bonaccini Calia, Domenico.
  Proceedings of the SPIE, Volume 6272, pp. 627210 (2006). Vol.~6272 of
  Presented at the Society of Photo-Optical Instrumentation Engineers (SPIE)
  Conference, {SPARTA: the ESO standard platform for adaptive optics real time
  applications}

\bibitem[\protect\citeauthoryear{Foley}{Foley}{2014}]{nvlink}
Foley D.,  2014, Technical report, NVLink, Pascal and Stacked Memory: Feeding
  the Appetite for Big Data,
  \verb+http://devblogs.nvidia.com/parallelforall/nvlink-pascal-stacked-memory-feeding-appetite-big-data+.
NVIDIA

\bibitem[\protect\citeauthoryear{{Foppiani}, {Diolaiti}, {Baruffolo},
  {Biliotti}, {Bregoli}, {Cosentino}, {Delabre}, {Lombini}, {Marchetti},
  {Rossettini}, {Schreiber}, {Tomelleri}, {Conan}, {D'Odorico} \&
  {Hubin}}{{Foppiani} et~al.}{2010}]{2010SPIE.7736E..99F}
{Foppiani} I.,  {Diolaiti} E.,  {Baruffolo} A.,  {Biliotti} V.,  {Bregoli} G.,
  {Cosentino} G.,  {Delabre} B.,  {Lombini} M.,  {Marchetti} E.,  {Rossettini}
  P.,  {Schreiber} L.,  {Tomelleri} R.,  {Conan} J.-M.,  {D'Odorico} S.,
  {Hubin} N.,  2010, in Society of Photo-Optical Instrumentation Engineers
  (SPIE) Conference Series Vol.~7736 of Society of Photo-Optical
  Instrumentation Engineers (SPIE) Conference Series, {System overview of the
  Multi conjugated Adaptive Optics RelaY for the E-ELT}

\bibitem[\protect\citeauthoryear{Gray \& Le~Roux}{Gray \&
  Le~Roux}{2012}]{efficientlqg}
Gray M.,  Le~Roux B., , 2012, Ensemble Transform Kalman Filter, a nonstationary
  control law for complex AO systems on ELTs: theoretical aspects and first
  simulations results

\bibitem[\protect\citeauthoryear{Hammer, Barbuy, Cuby, Kaper, Morris, Evans,
  Jagourel \& Puech}{Hammer et~al.}{2014}]{Hammer2014}
Hammer F.,  Barbuy B.,  Cuby J.,  Kaper L.,  Morris S.,  Evans C.,  Jagourel
  P.,    Puech M.,  2014, in Society of Photo-Optical Instrumentation Engineers
  (SPIE) Conference Series Vol. in print of Society of Photo-Optical
  Instrumentation Engineers (SPIE) Conference Series, {MOSAIC at E-ELT: a MOS
  for astrophysics, IGM, and cosmology}

\bibitem[\protect\citeauthoryear{Johns}{Johns}{2008}]{gmt}
Johns M.,  2008, in Extremely Large Telescopes: Which Wavelengths? Retirement
  Symposium for Arne Ardeberg Vol.~6986, The giant magellan telescope (gmt).
pp 698603--698603--12

\bibitem[\protect\citeauthoryear{McCalpin}{McCalpin}{1995}]{streambench}
McCalpin J.~D.,  1995, IEEE Computer Society Technical Committee on Computer
  Architecture (TCCA) Newsletter, 12, 19

\bibitem[\protect\citeauthoryear{{Myers}, {Hubert}, {Morris}, {Gendron},
  {Dipper}, {Kellerer}, {Goodsell}, {Rousset}, {Younger}, {Marteaud} \&
  {Basden}}{{Myers} et~al.}{2008}]{canaryshort}
{Myers} R.~M.,  {Hubert} Z.,  {Morris} T.~J.,  {Gendron} E.,  {Dipper} N.~A.,
  {Kellerer} A.,  {Goodsell} S.~J.,  {Rousset} G.,  {Younger} E.,  {Marteaud}
  M.,    {Basden} A.~G.,  2008, in Society of Photo-Optical Instrumentation
  Engineers (SPIE) Conference Series Vol.~7015 of Presented at the Society of
  Photo-Optical Instrumentation Engineers (SPIE) Conference, {CANARY: the
  on-sky NGS/LGS MOAO demonstrator for EAGLE}

\bibitem[\protect\citeauthoryear{{Nelson} \& {Sanders}}{{Nelson} \&
  {Sanders}}{2008}]{tmt}
{Nelson} J.,  {Sanders} G.~H.,  2008, in Society of Photo-Optical
  Instrumentation Engineers (SPIE) Conference Series Vol.~7012 of Society of
  Photo-Optical Instrumentation Engineers (SPIE) Conference Series, {The status
  of the Thirty Meter Telescope project}.
pp 70121A--70121A--18

\bibitem[\protect\citeauthoryear{{Rigaut}, {Neichel}, {Boccas}, {d'Orgeville},
  {Arriagada}, {Fesquet}, {Diggs}, {Marchant}, {Gausach}, {Rambold}, {Luhrs},
  {Walker}, {Carrasco-Damele}, {Edwards}, {Pessev} \& {Galvez}}{{Rigaut}
  et~al.}{2012}]{2012SPIE.8447E..0IRshort}
{Rigaut} F.,  {Neichel} B.,  {Boccas} M.,  {d'Orgeville} C.,  {Arriagada} G.,
  {Fesquet} V.,  {Diggs} S.~J.,  {Marchant} C.,  {Gausach} G.,  {Rambold}
  W.~N.,  {Luhrs} J.,  {Walker} S.,  {Carrasco-Damele} E.~R.,  {Edwards} M.~L.,
   {Pessev} P.,    {Galvez} R.~L.,  2012, in Society of Photo-Optical
  Instrumentation Engineers (SPIE) Conference Series Vol.~8447 of Society of
  Photo-Optical Instrumentation Engineers (SPIE) Conference Series, {GeMS:
  first on-sky results}

\bibitem[\protect\citeauthoryear{Sinharoy, Van~Norstrand, Eickemeyer, Le,
  Leenstra, Nguyen \& Konigsburg}{Sinharoy et~al.}{2015}]{power8short}
Sinharoy B.,  Van~Norstrand J.~A.,  Eickemeyer R.~J.,  Le H.~Q.,  Leenstra J.,
  Nguyen D.~Q.,    Konigsburg B.,  2015, IBM Journal of Research and
  Development, 59, 2:2

\bibitem[\protect\citeauthoryear{Sivo, Kulcsar, Conan, Raynaud, Gendron,
  Basden, Vidal \& Morris}{Sivo et~al.}{2014}]{lqgshort}
Sivo G.,  Kulcsar C.,  Conan J.,  Raynaud H.,  Gendron E.,  Basden A.,  Vidal
  F.,    Morris T.,  2014, Opt.\ Express

\bibitem[\protect\citeauthoryear{{Spyromilio}, {Comer{\'o}n}, {D'Odorico},
  {Kissler-Patig} \& {Gilmozzi}}{{Spyromilio} et~al.}{2008}]{eelt}
{Spyromilio} J.,  {Comer{\'o}n} F.,  {D'Odorico} S.,  {Kissler-Patig} M.,
  {Gilmozzi} R.,  2008, The Messenger, 133, 2

\bibitem[\protect\citeauthoryear{Starke, Stuecheli, Daly, Dodson, Auernhammer,
  Sagmeister, Guthrie, Marino, Siegel \& Blaner}{Starke
  et~al.}{2015}]{power8memory}
Starke W.~J.,  Stuecheli J.,  Daly D.,  Dodson J.,  Auernhammer F.,  Sagmeister
  P.~M.,  Guthrie G.~L.,  Marino C.~F.,  Siegel M.,    Blaner B.,  2015, IBM
  Journal of Research and Development, 59(1), 3:1

\bibitem[\protect\citeauthoryear{Stuecheli, Blaner, Johns \& Siegel}{Stuecheli
  et~al.}{2015}]{capi}
Stuecheli J.,  Blaner B.,  Johns C.,    Siegel M.,  2015, IBM Journal of
  Research and Development, 59, 7:1

\bibitem[\protect\citeauthoryear{Vernet, Cayrel, Hubin, Mueller, Biasi,
  Gallieni \& Tintori}{Vernet et~al.}{2012}]{30year}
Vernet E.,  Cayrel M.,  Hubin N.,  Mueller M.,  Biasi R.,  Gallieni D.,
  Tintori M., , 2012, Specifications and design of the E-ELT M4 adaptive unit

\end{thebibliography}
\bsp

\end{document}